\documentclass[12pt]{article}

\usepackage{feynmp}                       
\newcommand{\bib}[1]{\bibitem{#1}}

\newcommand{\3}{\ss}
\newcommand{\ia}{{\"{\i}}}  
\newcommand{\absatz}{\vspace{2ex}\noindent}

\newcommand{\journal}[4]{{#1} {\bf{#2}}, #3 (#4)}

\newcommand{\NPB}{\emph{Nucl.\ Phys.\ }{B}}
\newcommand{\PLB}{\emph{Phys.\  Lett.\ }{B}}
\newcommand{\PRD}{\emph{Phys.\ Rev.\ }{D}}

\newcommand{\preprint}[4]{\emph{#2}; #3, #4 #1}

\newcommand{\non}{\nonumber}

\newcommand{\ii}{\mathrm{i}}
\newcommand{\dd}{\mathrm{d}}

\newcommand{\xs}{\vec{X}_{\mathrm{s}}}
\newcommand{\ts}{T_{\mathrm{s}}}

\newcommand{\xu}{\vec{X}_{\mathrm{u}}}
\newcommand{\tu}{T_{\mathrm{u}}}
\newcommand{\Qs}{Q_{\mathrm{s}}}
\newcommand{\Qp}{Q_{\mathrm{p}}}

\newcommand{\Ap}{A_{\mathrm{p}}}

\newcommand{\Amu}{{A^\mu}}
\newcommand{\Asmu}{{A_{{\rm s}}^\mu}}
\newcommand{\Apmu}{{A_{{\rm p}}^\mu}}
\newcommand{\Aumu}{{A_{{\rm u}}^\mu}}

\newcommand{\deintdim}[2]{\frac{\dd^{#1}\! #2}{(2\pi)^{#1}}\;}

\newcommand{\kv}{\vec{k}}
\newcommand{\pv}{\vec{\,\!p}\!\:{}}

\newcommand{\xv}{\vec{x}}

\begin{document}
\begin{fmffile}{mcebfeyn}
  \fmfset{curly_len}{2mm} \fmfset{wiggly_len}{3mm}
  \newcommand{\feynbox}[2]{\mbox{\parbox{#1}{#2}}}
\newcommand{\fs}{\scriptstyle} 
\newcommand{\hq}{\hspace{0.5em}}
\begin{titlepage}
\begin{flushright}
  hep-ph/9810236\\ NT@UW-98-23 \\ 1st October 1998 \\
\end{flushright}
\vspace*{1.5cm}
\begin{center}
  
  \LARGE{\textbf{The Soft R\'egime and $\beta$ Function of NRQCD}}

\end{center}
\vspace*{1.0cm}
\begin{center}
  \textbf{Harald W.\ Grie\3hammer\footnote{Talk presented at the
      ``Euroconference QCD '98'' in Montpellier, France, 2nd -- 8th July 1998
      (to appear in \NPB (Proc.\ Suppl.)), and at the conference ``Quark
      Confinement and the Hadron Spectrum III'' at TJNL, Newport News, USA, 8th
      -- 12th June 1998 (to appear in the proceedings). Email:
      hgrie@phys.washington.edu}}
  
  \vspace*{0.2cm}
  
 \emph{Nuclear Theory Group, Department of Physics, University of Washington,\\
    Box 351 560, Seattle, WA 98195-1560, USA} \vspace*{0.2cm}

\end{center}

\vspace*{2.0cm}

\begin{abstract}
  Progress towards a complete velocity power counting in non-relativistic
  effective field theories, especially NRQCD, is motivated and summarised.
\end{abstract}
\vspace{0.7cm}

\end{titlepage}

\noindent
In Non-Relativistic QCD~\cite{CaswellLepage,BBL}{}, a heavy quark is nearly on
shell and its energy is dominated by its mass $M$, so that the resulting
non-relativistic system can be treated perturbatively both in the coupling
constant $g$ and in the quark velocity $v$. The typical energy and momentum
scales are the non-relativistic kinetic energy $Mv^2$ and the momentum $Mv$.
The effective Lagrangean does not exhibit $v$ explicitly, so that a power
counting scheme has to be established which determines uniquely which terms in
the Lagrangean must be taken into account to render predictive power to a given
precision in $v$.

Identification of the relevant energy and momentum r\'egimes has proven
difficult: The first attempt by Lepage and co-workers~\cite{LMNMH} fell shot as
working only in the Coulomb gauge and as being incomplete. In a recent article,
Beneke and Smirnov~\cite{BenekeSmirnov} pointed out that the much simpler
velocity rescaling rules proposed by Luke and Manohar for Coulomb
interactions~\cite{LukeManohar}{}, and by Grinstein and Rothstein for
bremsstrahlung processes~\cite{GrinsteinRothstein}{}, as united by Luke and
Savage~\cite{LukeSavage}{}, and by Labelle's power counting scheme in time
ordered perturbation theory~\cite{Labelle}{}, do not reproduce the two gluon
exchange between particles near threshold.  Recently, velocity power counting
rules using dimensional regularisation were established for a toy
model~\cite{hgpub3} and NRQCD~\cite{hgpub5} (see also~\cite{hgpub4}) following
Beneke and Smirnov's threshold expansion~\cite{BenekeSmirnov}{}. Here, I
summarise two recent publications~\cite{hgpub3,hgpub5}{}.

\absatz The NRQCD propagators are read off from the Lagrangean as
\begin{equation}
  \label{nonrelprop}
  Q\;:\;\frac{\ii\;\mathrm{Num}}{T-\frac{\pv^2}{2M}+\ii\epsilon}\;\;,\;\;
   \Amu\;:\;\frac{\ii\;\mathrm{Num}}{k^2+\ii\epsilon}\;\;,
\end{equation}
where $T=\frac{\pv^2}{2M}+\dots$ is the kinetic energy of the quark.
``$\mathrm{Num}$'' are numerators containing the appropriate colour, Dirac and
flavour indices. Cuts and poles in scattering amplitudes stem from bound states
and on-shell propagation of particles in intermediate states and give rise to
infrared divergences. With energies and momenta of either scale, there are only
three r\'egimes in which quarks or gluons in (\ref{nonrelprop}) are on shell:
\begin{eqnarray}
   \mbox{soft r\'egime: }&\Asmu:&\;k_0\sim |\kv|\sim Mv\;\;,\non\\
  \label{regimes}
   \mbox{potential r\'egime: }&Q_\mathrm{p}:&\;T\sim Mv^2\;,\; |\pv|\sim
   Mv\;\;,\\
   \mbox{ultrasoft r\'egime: }&\Aumu:&\;k_0\sim |\kv|\sim Mv^2\non
\end{eqnarray}
Ultrasoft gluons $\Aumu$ are emitted as bremsstrahlung or from excited states
in the bound system, and hence physical.  Overall energy conservation forbids
all processes with outgoing soft gluons but without ingoing ones.  Gluons which
change the quark momenta but keep them close to their mass shell relate the
(instantaneous) Coulomb interaction:
\begin{equation}
  \label{pgluon}
  \Apmu\;\;:\;\;k_0\sim Mv^2\;,\;|\kv|\sim Mv
\end{equation}
When a soft gluon $\Asmu$ couples to a potential quark $ Q_\mathrm{p}$, the
outgoing quark is far off its mass shell and carries energy and momentum of
order $Mv$.
\begin{equation}
  \label{squark}
  Q_\mathrm{s}\;\;:\;\;T\sim |\pv|\sim Mv
\end{equation}
As the potential quark has a much smaller energy than either of the soft
particles, it can -- by the uncertainty relation -- not resolve the precise
time at which the soft quark emits the soft gluon. So, a ``temporal'' multipole
expansion is associated with this vertex. In general, the coupling between
particles of different r\'egimes will not be point-like but contain multipole
expansions for the particle belonging to the weaker kinematic r\'egime.

One may expand~\cite{BenekeSmirnov} the integrand of a loop integral in NRQCD
about the various saddle points, i.e.\ about the values of the loop-momentum
$q$ where particles become on shell. For example, expanding about a physical
gluon at the soft scale, the quark propagator is ($T_\mathrm{p}\sim
\frac{\vec{p}^2}{2M}\sim Mv^2\ll q_0\sim|\vec{q}|\sim Mv$)
\begin{equation}
  \label{bsexprop}
   \frac{\ii}{{q_{0,\rm s}}+{T_\mathrm{p}}-\frac{(\vec{p}+\vec{q})^2}{2M}}
   \longrightarrow
   \frac{\ii}{{q_{0,\rm s}}}+\frac{\ii}{{q_{0,\rm s}}}\;\ii\bigg(
     {T_\mathrm{p}}-\frac{(\vec{p}+\vec{q})^2}{2M}\bigg)\;
     \frac{\ii}{{q_{0,\rm s}}}+\dots\;\;.
\end{equation}
Higher order terms in the expansion are interpreted as insertions into the soft
quark propagator (static to lowest order) and as multipole expansions which
modify the vertex rules.  As the energy of potential gluons is much smaller
than their momentum, the $\Ap$-propagator becomes instantaneous.  With these
five fields $\Qs,\Qp,\Asmu,\Apmu,\Aumu$ representing the relevant
infrared-relevant degrees of freedom, i.e.\ quarks and gluons in the three
different non-relativistic r\'egimes, soft, potential and ultrasoft, NRQCD
becomes self-consistent.

The quark propagator is gauge independent. Gauge fixing introduces gauge
dependent denominators multiplying the gauge independent denominators in the
gluon propagators. When they are non-zero for all but the above combinations of
scales $Mv$ and $Mv^2$ (\ref{regimes}), the decomposition into the three
r\'egimes (\ref{regimes}) remains unchanged, as do the rescaling properties of
the fields and interactions. The standard gauges (axial, Weyl, Lorentz,
Coulomb) will therefore all show the same power counting and vertex rules.
Details of the gluon propagator and its insertions are different in different
r\'egimes and gauges, and some gauges will not exhibit certain vertices,
insertions and representatives; e.g.\ the Coulomb gauge is unique in having
$A_0$ contribute only in the potential r\'egime as physical fields are
transverse by virtue of Gau\3' law.

Finally, the regularisation scheme must be chosen such that expansion around
one saddle point in the loop integral does not obtain any contribution from
other saddle points. Cutoff regularisation jeopardises power counting and
symmetries, and introduces unphysical power divergences. In contradistinction,
using dimensional regularisation \emph{after} the saddle point expansion
preserves power counting and gauge symmetry. Its homogeneity guarantees that
contributions from different saddle points and r\'egimes do not overlap. (A
simple example is given in Ref.\ \cite{hgpub3}{}.)

In order to establish explicit $v$ counting in the NRQCD Lagrangean, one
rescales the space-time coordinates such that typical momenta in each r\'egime
are dimensionless:
\begin{eqnarray}
  \mbox{soft: } && t=(Mv)^{-1} \;\ts\;\;,\;\; \xv=(Mv)^{-1}\;\xs\;\;,\non\\
  \label{xtscaling}
  \mbox{potential: }&& t=(Mv^2)^{-1}\; \tu\;\;,\;\; \xv=(Mv)^{-1}\;\xs\;\;,\\
\mbox{ultrasoft: }&& t=(Mv^2)^{-1}\; \tu\;\;,\;\;\xv=(Mv^2)^{-1}\;\xu\;\;.\non
\end{eqnarray}
The scaling of the fields is derived by normalising the propagator terms to be
of order $v^0$. In order to maintain velocity power counting, corrections of
order $v$ or higher in the thus rescaled Lagrangean must be treated as
insertions.
Particles in the various r\'egimes couple: On-shell (potential) quarks radiate
bremsstrahlung (ultrasoft) gluons. In general, one must allow all couplings
between the various r\'egimes which obey ``scale conservation'' for energies
and momenta. This excludes for example the coupling of two potential quarks
($T\sim Mv^2$) to one soft gluon ($q_0\sim Mv$), but not to two soft gluons.

Amongst the fields introduced, two (six) interactions are allowed for example
within and between the various r\'egimes for the scalar coupling in the Coulomb
(Lorentz) gauge whose $v$ counting is read up from the rescaled
Lagrangean~\cite{hgpub3,hgpub5,hgpub4}{}, see table \ref{scalarvertex}. Albeit
both describing interactions with physical gluons, soft and ultrasoft couplings
occur at different orders in $v$, and obey different multipole expansion rules.
Double counting is prevented because in addition to most of the propagators,
all vertices involve different multipole expansions. Because the vertex rules
for the soft r\'egime count powers of $v$ with respect to the soft r\'egime,
one retrieves there the velocity power counting of Heavy Quark Effective
Theory~\cite{IsgurWise1,IsgurWise2}{}. HQET becomes a sub-set of NRQCD,
complemented by interactions between soft (HQET) and potential or ultrasoft
particles. 
\begin{table}[!htb]
  \caption{\label{scalarvertex}\sl Velocity power counting and vertices for the
     interaction Lagrangean \protect$-g Q^\dagger A_0 Q$ in the Lorentz gauge.
     In the Coulomb gauge, only the first two diagrams exist.
     Zigzagged, dashed and wiggly lines represent \protect$\Asmu$,
     \protect$\Apmu$ and \protect$\Aumu$, double and single lines
     \protect$\Qs$ and \protect$\Qp$, respectively.}
\begin{center}
  \footnotesize
\begin{tabular}{|r||c|c||c|c|c|c|}
  \hline
  Vertex\rule[-17pt]{0pt}{40pt}&
      \feynbox{40\unitlength}{
      \begin{fmfgraph*}(40,25)
        \fmfstraight \fmftop{i,o} \fmfbottom{u} \fmf{vanilla,width=thick}{i,v}
        \fmf{vanilla,width=thick}{o,v} \fmffreeze
        \fmf{dashes,width=thin,label=$\fs A_0$}{u,v}
      \end{fmfgraph*}}&
      \feynbox{40\unitlength}{
        \begin{fmfgraph*}(40,25)
          \fmfstraight \fmftop{i,o} \fmfbottom{u} \fmf{double,width=thin}{i,v}
          \fmf{double,width=thin}{o,v} \fmffreeze
          \fmf{dashes,width=thin,label=$\fs A_0$}{u,v}
      \end{fmfgraph*}}&
      \feynbox{40\unitlength}{
      \begin{fmfgraph*}(40,25)
        \fmfstraight \fmftop{i,o} \fmfbottom{u} \fmf{vanilla,width=thick}{i,v}
        \fmf{vanilla,width=thick}{o,v} \fmffreeze \fmf{photon,width=thin}{u,v}
      \end{fmfgraph*}}&
      \feynbox{40\unitlength}{
      \begin{fmfgraph*}(40,25)
        \fmfstraight \fmftop{i,o} \fmfbottom{u} \fmf{double,width=thin}{i,v}
        \fmf{double,width=thin}{o,v} \fmffreeze \fmf{zigzag,width=thin}{u,v}
      \end{fmfgraph*}}&
      \feynbox{40\unitlength}{
      \begin{fmfgraph*}(40,25)
        \fmfstraight \fmftop{i,o} \fmfbottom{u} \fmf{double,width=thin}{i,v}
        \fmf{vanilla,width=thick}{o,v} \fmffreeze \fmf{zigzag,width=thin}{u,v}
      \end{fmfgraph*}}&
      \feynbox{40\unitlength}{
      \begin{fmfgraph*}(40,25)
        \fmfstraight \fmftop{i,o} \fmfbottom{u} \fmf{double,width=thin}{i,v}
        \fmf{double,width=thin}{o,v} \fmffreeze \fmf{photon,width=thin}{u,v}
      \end{fmfgraph*}}
  \\[2ex]
  \hline
  \protect$v$ power\rule[-10pt]{0pt}{24pt}&
  \protect$\frac{1}{\sqrt{v}}$&
  \protect$\sqrt{v}$&
  \protect$v^0$&
  \protect$v^0$&
  \protect$v^0$&
  \protect$v$
  \\
  \hline
\end{tabular}
\end{center}
\end{table}
 In NRQCD with two potential quarks as initial and final states, the
power counting in a soft sub-graph has to be transfered to the potential
r\'egime~\cite{hgpub3,hgpub5}{}.  Because soft loop momenta scale like
$[\dd^{4}\!  k_\mathrm{s}]\sim v^4$, while potential ones like $[\dd^{4}\!
k_\mathrm{p}]\sim v^5$, each largest sub-graph which contains soft particles is
enhanced by an additional factor $\frac{1}{v}$.

\absatz Matching between NRQCD and QCD proved that the soft quark and gluon are
indispensable to reproduce the correct structure of collinear (infrared)
divergences of the two gluon exchange contribution to Coulomb scattering
between non-relativistic particles near threshold in a toy model and confirmed
the proposed counting rules~\cite{hgpub3}{}.

Because NRQCD is a well-defined field theory of quarks and gluons, one can
investigate its renormalisation group equations under the assumption that
perturbation theory is applicable at all scales. NRQCD must reproduce the QCD
$\beta$ function with $N_\mathrm{F}$ light quarks below the scale $M$. The
result thus already anticipated,
\begin{equation}
  \label{qcdbetafunction}
  \beta_\mathrm{NRQCD}=-\;\frac{g_\mathrm{R}^3}{(4\pi)^2}\;
     \left[\frac{11}{3}\;N-\frac{2}{3}\;N_\mathrm{F}\right]
\end{equation}
to lowest order for the gauge group SU($N$), the prime goal of this
exercise~\cite{hgpub5} is not the result but to provide further insight: The
Lorentz gauge is a legitimate gauge choice and simplifies renormalisation. The
renormalised coupling strengths of all interactions steming from expanding the
same term in the Lagrangean in the various r\'egimes are the same, except that
they have to be taken at different scales. Although the number of vertices is
increased, the number of independent couplings is not. The soft r\'egime is
indispensable for NRQCD to describe especially the correct running coupling for
the Coulomb gluons, as their sole contribution in the vacuum polarisation comes
from the propagation of soft (on-shell) gluons in the loop. The computation
also confirms the validity of the power counting proposed and is slightly
simpler than its QCD counterparts.  The non-Abelian nature of the gauge field
enters to lowest order only via the non-zero contribution of the three gluon
vertex to the vacuum polarisation. Non-Abelian vertex contributions are
down by at least one power in $v$ compared to  Abelian vertex corrections
and bare vertices. The quark self energy and vertex renormalisations differ
from its QCD counterpart, but the Slavnov--Taylor identities hold.

From a technical point of view, the number of possible diagrams in power
counted NRQCD is considerably larger than in the original theory because there
are at least six vertices per interaction allowed by scale conservation, cf.\ 
table \ref{scalarvertex}.  The calculation is greatly simplified by
diagrammatic rules~\cite{hgpub5} which follow from the homogene{\ia}ty property
of dimensional regularisation. That integrals without scales set by external
momenta or energies in the denominator vanish in dimensional regularisation,
\begin{equation}
  \label{masterformula}
  \int\deintdim{d}{q} q^\alpha=0\;\;,
\end{equation}
allows one to recognise the majority of NRQCD (and threshold
expansion~\cite{BenekeSmirnov}{}) graphs as zero to all orders in $v$ just by
drawing them, independently of details of the vertices involved. The concept is
sensitive only to the multipole expansion of the vertices and independent of
insertions and the chosen gauge. Only the denominators of the graph lowest
order in $v$ have to be looked at.  It is usually only necessary to consider a
sub-set of the whole diagram. One assigns the loop and external momenta to it,
taking into account multipole expansions, and writes out the denominators
(i.e.\ inverse propagators). If by shifting integration variables, one arrives
at an energy or momentum integral without scale, the diagram vanishes, as do
any diagrams which contain it as a sub-graph. As an example, the following
rule may be quoted, which states that a potential gluon coupled to two soft
particles, quark or gluon, vanishes because there is no scale in the $q_0$
part of the loop integration, even when potential or ultrasoft particles
couple to the Coulomb gluon:
\vspace*{-1.5ex}
\begin{equation}
  \hq\hq\hq\hq\
  \feynbox{70\unitlength}{
      \begin{fmfgraph*}(70,50)
        \fmftop{t0,t01,t1,t2,t3,t4,t41}
        \fmfleft{bl,tl}
        \fmfright{br,tr}
        \fmf{zigzag}{tl,vm1,bl}
        \fmf{double}{tl,vm1,bl}
        \fmf{zigzag}{tr,vm2,br}
        \fmf{double}{tr,vm2,br}
        \fmffreeze
        \fmf{dashes}{vm1,v1,vm2}
        \fmfblob{10}{v1}
        \fmffreeze
        \fmf{dashes}{v1,v2}
        \fmf{dashes,tension=4}{v2,t1}
        \fmf{photon}{v1,v3}
        \fmf{photon,tension=4}{v3,t2}
        \fmf{vanilla}{v1,v4}
        \fmf{vanilla,tension=4}{v4,t3}
        \fmffreeze
        \fmf{dots}{v2,v3,v4}
      \end{fmfgraph*}}
    \hq\hq=0
\end{equation}
In the case of the NRQCD $\beta$ function, the number of vertices for each of
the couplings of table \ref{scalarvertex} is reduced from three or more to
one, and the number of vacuum polarisations also from three to one. This
result holds for any gauge and any interaction involving loops with three
particles, showing the general usefulness of these rules.

For a derailed account of the points outlined in this contribution, I refer
to~\cite{hgpub3,hgpub5} and references therein. In conclusion, a complete
velocity power counting of NRQCD has been suggested. The results are readily
generalised to any effective field theory with two or more infrared scales.

\vskip 0.7 cm \thebibliography{References}

\bib{CaswellLepage} W.\ E.\ Caswell and G.\ P.\ Lepage:
\journal{\PLB}{167}{437}{1986}.
  
\bib{BBL} G.\ T.\ Bodwin, E.\ Braaten and G.\ P.\ Lepage:
\journal{\PRD}{51}{1125}{1995}; \journal{\PRD}{55}{5853}{1997}.

\bib{LMNMH} G.\ P.\ Lepage et al.: \journal{\PRD}{46}{4052}{1992}.

\bib{BenekeSmirnov} M.\ Beneke and V.\ A.\ Smirnov: \preprint{}{Asymptotic
  Expansion of Feynman Integrals near Threshold}{CERN-TH/7-315,
  hep-ph/9711391}{1997}.

\bib{LukeManohar} M.\ Luke and A.\ V.\ Manohar: \journal{\PRD}{55}{4129}{1997}.

\bib{GrinsteinRothstein} B.\ Grinstein and I.\ Z.\ Rothstein:
\journal{\PRD}{57}{78}{1998}.

\bib{LukeSavage} M.\ Luke and M.\ J. Savage: \journal{\PRD}{57}{413}{1998}.

\bib{Labelle} P.\ Labelle: \preprint{}{Effective Field Theories for QED Bound
  States: Extending Nonrelativistic QED to Study Retardation
  Effects}{MCGILL-96-33, hep-ph/9608491}{1996}.

\bib{hgpub3} H.\ W.\ Grie\3hammer, \preprint{(to be published in \PRD $\;$ {\bf
     58})}{Threshold Expansion and Dimensionally Regularised NRQCD}{NT@UW-98-3,
  hep-ph/9712467}{1997}.

\bib{hgpub5} H.\ W.\ Grie\3hammer, \preprint{}{Power Counting and $\beta$
  Function in NRQCD}{NT@UW-98-22, hep-ph/9810235}{1998}.
  
\bib{hgpub4} H.\ W.\ Grie\3hammer, \preprint{(Proceedings of the Workshop
  ``Nuclear Physics With Effective Field Theories'' at Caltech, 26th -- 27th
  February 1998, eds.\ R.\ Seki, U.\ van Kolck and M.\ J.\ Savage, World
  Scientific)}{The Soft R\'egime in NRQCD}{NT@UW-98-12, hep-ph/9804251}{1998}.

\bib{IsgurWise1} N.\ Isgur and M.\ B.\ Wise: \journal{\PLB}{232}{113}{1989}.

\bib{IsgurWise2} N.\ Isgur and M.\ B.\ Wise: \journal{\PLB}{237}{527}{1990}.
  
\end{fmffile}
\end{document}